\documentclass[
 reprint,
 amsmath,amssymb,
 aps, groupedaddress
]{revtex4-2}

\usepackage{graphicx}
\usepackage{dcolumn}
\usepackage{bm}
\usepackage{xcolor}
\usepackage{braket}
\usepackage{mathtools}
\usepackage{hyperref}

\begin{document}

\preprint{APS/123-QED}

\title{Toward Resilient Many-Body Formulations under Incomplete Correlation Models: A Dual-Space Variational Formulation}

\author{Vibin Abraham}
\affiliation{Integrated Discovery Sciences Directorate, Pacific Northwest National Laboratory, Richland, Washington 99354, USA}
\author{Bhumika Jayee}
\affiliation{Integrated Discovery Sciences Directorate, Pacific Northwest National Laboratory, Richland, Washington 99354, USA}
\author{Bo Peng}
\affiliation{Integrated Discovery Sciences Directorate, Pacific Northwest National Laboratory, Richland, Washington 99354, USA}
\author{Nicholas P. Bauman}
\affiliation{Integrated Discovery Sciences Directorate, Pacific Northwest National Laboratory, Richland, Washington 99354, USA}
\author{Karol Kowalski}
\email{karol.kowalski@pnnl.gov}
\affiliation{Integrated Discovery Sciences Directorate, Pacific Northwest National Laboratory, Richland, Washington 99354, USA}
\affiliation{Department of Physics, University of Washington, Seattle, Washington 98195, USA}

\date{\today}

\begin{abstract}
Quantum simulations of correlated many-body systems will inevitably operate with imperfect information: the prepared states may arise from incomplete or approximate ansätze, while their measured Hamiltonian and overlap matrix elements are additionally affected by hardware and statistical noise. 
Robust quantum algorithms will be those that can improve upon reliable lower-level descriptions in spite of these imperfections.
We introduce a dual-space nonorthogonal configuration-interaction (DS-NOCI) framework built around this principle. Rather than replacing a classically accessible NOCI reference manifold with correlated quantum states, DS-NOCI retains both spaces and couples them variationally. The reference sector provides a stable many-body backbone, while the quantum states contribute whatever additional correlation directions they contain.
With exact matrix elements, the enlarged dual space gives a ground-state energy no higher than either the reference-only NOCI or correlated-only quantum-NOCI spaces, ensuring that an incomplete correlation model cannot degrade the underlying variational description. The additional reference--correlated matrix elements require only one correlated state preparation and are therefore less demanding than the correlated--correlated block already required by quantum variant of NOCI. Molecular benchmarks further show enhanced resilience to imperfect amplitudes and noisy matrix elements. DS-NOCI thus provides a general strategy for building quantum many-body formulations that remain useful when both correlation models and quantum computations are necessarily incomplete and noisy.
\end{abstract}

\maketitle

\section{Introduction}

Accurate electronic-structure calculations are central to predictive
chemistry and materials science, but strongly correlated problems remain
challenging because many electronic configurations can become simultaneously
important. Quantum computers offer a different representation of the
many-electron problem and have motivated algorithms intended to access
molecular energies and wave functions beyond the practical reach of direct
classical  diagonalization~\cite{reiher2017elucidating,cao2019quantum,mcardle2020quantum,bauer2020quantum,daley2022practical}.

Two familiar endpoints are quantum phase estimation and variational quantum
eigensolvers (VQEs)
\cite{peruzzo2014vqe,mcclean2016theory,romero2018strategies,Kandala2017,Kandala2019,izmaylov2019unitary,lang2020unitary,verteletskyi2020measurement,grimsley2019trotterized,mcardle2020quantum,Love2021,tilly2022variational}.
Phase-estimation-based approaches 
\cite{nielsen2002quantum,kitaev1995quantum,Kitaev_1997,Abrams1999QPE,childs2010relationship,reiher2017elucidating}
offer a systematic
route to eigenvalues but demand long coherent evolutions and sufficiently
good initial-state overlap. VQE reduces coherence requirements by preparing
a parameterized state and minimizing its measured energy in a
quantum--classical loop~\cite{peruzzo2014vqe,mcclean2016theory,cerezo2021vqaReview}.
The latter strategy replaces some coherent depth by nonlinear optimization,
whose performance can depend strongly on the ansatz, parameterization, and
optimization landscape; barren plateaus are one well-known manifestation of
this difficulty~\cite{mcclean2018barren}.

Quantum subspace methods provide a broad alternative. Instead of asking a
single parameterized circuit to represent an eigenstate, they prepare a
collection of states, measure Hamiltonian and overlap matrix elements among
them, and solve the resulting projected generalized eigenvalue problem.
This viewpoint encompasses quantum subspace expansion~\cite{mcclean2017hybrid},
quantum imaginary-time and Lanczos constructions~\cite{motta2020qite},
multireference and other quantum Krylov methods~\cite{stair2020mrqk,cortes2022quantum},
quantum filter diagonalization~\cite{parrish2019qfd}, and a growing collection
of related methods reviewed in Ref.~\cite{motta2024subspace}. These methods
trade a global nonlinear search for the design, preparation, measurement, and
conditioning of a compact subspace. Nonorthogonality is not incidental in
this setting: it permits physically distinct or differently generated states
to be combined without forcing them into a common orthogonal
parameterization, although the resulting overlap metric can introduce
numerical sensitivity~\cite{epperly2022qsd}. In multireference Krylov
constructions, the original reference states can appear naturally as the
zero-time members of the generated Krylov spaces~\cite{stair2020mrqk};
thus, the parent multireference information is retained by construction when
those vectors are included.

Several approaches exploit nonorthogonal quantum subspaces more explicitly.
The nonorthogonal variational quantum eigensolver (NOVQE) diagonalizes in a
space of multiple independently parameterized quantum
states~\cite{huggins2020nonorthogonal}. The nonorthogonal quantum eigensolver
(NOQE) provides a particularly relevant connection to quantum chemistry:
classically optimized symmetry-broken Hartree--Fock determinants are used as
references for reference-specific correlated quantum states, whose
Hamiltonian and overlap matrix elements define a generalized eigenvalue
problem~\cite{baek2023noqe}. Generator-coordinate-inspired methods (GCIM) similarly
construct overcomplete nonorthogonal many-body bases from excitation
generators~\cite{zheng2023quantum,zheng2024gcim}. Subsequent developments of NOQE have focused
on reducing measurement cost and improving robustness through, for example,
shadow-based estimation and amplitude-estimation
strategies~\cite{ren2025shadownoqe,ren2026measurementnoqe}. Collectively,
these methods establish nonorthogonal subspace diagonalization as a viable
architecture for quantum eigensolvers.

There is a closely related and much longer history of developing configuration interaction algorithms for classical computers.
Nonorthogonal configuration interaction (NOCI) mixes distinct
Hartree--Fock or otherwise constructed reference determinants and can provide
compact descriptions of static correlation~\cite{thom2009hartreefock}.
Subsequent work has explored multielectron excited states and alternative
strategies for constructing useful NOCI manifolds, including spin-flip and
SCF-state-based approaches~\cite{sundstrom2014noci,mayhall2014sfnoci,
burton2019noci}. NOCI ideas have also been extended to core-excited and
core-ionized states~\cite{oosterbaan2018nocis,hait2020nocis}, fragment-based
descriptions~\cite{sanchezmansilla2022nocif}, and size-consistent
constructions based on localized spin rotations~\cite{lee2022localized}.
Dynamic-correlation corrections have been introduced through perturbative
NOCI approaches~\cite{burton2020nocipt2}, while selected NOCISD and
natural-orbital NOCI seek compact and more systematic nonorthogonal
determinant expansions~\cite{sun2024snocisd,graf2024nonoci}. Recent methods
such as EIDOS and its excited-state extension EXIDOS further demonstrate that
highly accurate many-electron states can be represented using relatively
small numbers of independently optimized nonorthogonal Slater
determinants~\cite{giuliani2026eidos,exidos2026}.

These classical and quantum developments raise a simple structural question.
Suppose a reliable classical NOCI calculation has already identified a useful
nonorthogonal reference manifold, and each reference is subsequently used to
generate a correlated quantum state. A natural extension, as in NOQE-like
constructions, is to diagonalize in the space formed by the correlated
states. The correlated states, however, need not reproduce the original
reference manifold exactly, and the parent NOCI states need not remain
available as independent variational degrees of freedom. Replacing the
reference manifold by its correlated descendants can therefore discard
variational flexibility that was already present in the classical NOCI
description. This issue is distinct from the quality of the correlated-state
generator itself.

The observation suggests a minimal alternative: if quantum correlation is
intended to augment a useful classical NOCI description, retain the
classical reference states explicitly while adding the quantum-generated
states. This retention occurs automatically in some Krylov constructions,
where the starting states are members of the generated span, but is not a
general property of reference-specific correlated-state
approaches~\cite{stair2020mrqk,baek2023noqe}. In this sense, the present work
does not require a new correlated-state generator. Rather, it concerns how
the classical reference information is carried into the final quantum
subspace calculation. NOQE provides a natural realization of this idea:
its reference-specific correlated states can be supplemented by the
underlying NOCI references without changing the machinery used to generate
those correlated states.

This viewpoint is also consistent with a broader trend toward
quantum--classical partitioning in electronic-structure theory. Classical
electronic-structure information can be used to construct compact
multideterminantal states for quantum computation or to restrict the quantum
calculation to selected configuration spaces~\cite{fomichev2024stateprep,
aydogan2026ssvqci}. Multireference unitary coupled-cluster approaches
similarly use multiple important configurations to construct correlated
quantum ans"atze~\cite{greenediniz2021mrucc,wu2025mrucc,chawla2025mrucc}.
Conversely, quantum-computed active-space or multireference information can
be passed to classical dynamic-correlation treatments, including tailored
and externally corrected coupled cluster, multireference perturbation
theory, and auxiliary-field quantum Monte
Carlo~\cite{erhart2024tcc,scheurer2024ecc,krompiec2023nevpt2,
erhart2025qscitcc,yoshida2025qsciafqmc}. These approaches differ
substantially in implementation, but share the principle that quantum
resources are most useful when embedded within, rather than isolated from,
the chemical information and correlation machinery available classically.

Quantum-selected configuration interaction (QSCI) provides a complementary
example of this division of labor. QSCI uses quantum sampling to identify
important electronic configurations but constructs and diagonalizes the
Hamiltonian in the resulting determinant space
classically~\cite{kanno2023qsci,robledo2025chemistry}. Its usefulness therefore remains tied to
the ability to obtain a compact selected space: increasing the number of
retained determinants ultimately enlarges the same type of classical CI
diagonalization that selected-CI approaches seek to control. Recent analyses
have highlighted limitations associated with repeated sampling and the
compactness of the resulting determinant spaces relative to established
classical selected-CI strategies~\cite{reinholdt2025qsci}. Conversely,
subspace-selected variational QCI transfers a classically selected
configuration space to a quantum variational representation specifically to
avoid a large classical matrix diagonalization~\cite{aydogan2026ssvqci}.
Very recent work has also combined NOCI-optimized orbital bases with quantum
sampling, showing that classical nonorthogonal representations can improve
the discovery of useful configurations under a fixed classical
diagonalization budget~\cite{vanrossum2026nociqsci}. These developments
further motivate designing quantum algorithms around compact, chemically
informed subspaces rather than treating classical and quantum descriptions
as competing alternatives.

We therefore formulate \emph{Dual-Space Nonorthogonal Configuration
Interaction} (DS-NOCI) by retaining the full parent reference manifold as
one variational sector and coupling it to the generated correlated manifold
as a second sector. The resulting variational space contains both the
classical NOCI references and their correlated companions. The construction
may therefore be viewed as an augmentation of NOCI, NOQE, or another
reference-specific state-generation procedure rather than as a replacement
for those methods. Its defining feature is that both the classical reference
information and the additional correlated directions remain available to the
final Rayleigh--Ritz problem.

This construction has three immediate consequences. First, because the
dual space contains both parent spaces, its exact-matrix Rayleigh--Ritz
ground-state energy cannot be higher than either the reference-only or
correlated-only result. For a NOQE-type application, for example, the final
space contains both the original classical NOCI manifold and the correlated
NOQE manifold. Second, changes to a generated state that remain entirely
within the retained reference manifold alter its representation but do not
introduce a new variational direction. The genuinely new information is
therefore carried only by components generated outside the reference
manifold. Third, the DS construction is independent of the mechanism used
to generate those directions. Fixed perturbative amplitudes, partially
optimized circuits, fully optimized correlated states, Krylov propagation,
or other future state-generation procedures can all be used within the same
dual-space framework.

The numerical calculations below are deliberately small proof-of-principle
tests. They use inexpensive reference-specific state generators to isolate
the effect of retaining the parent space, and they include controlled tests
of state-parameter and matrix-element errors. The goal is not to claim that
a particular perturbative generator is universally optimal, nor to replace
existing NOCI or NOQE constructions, but to demonstrate a general principle:
when correlated quantum states are generated from an already useful
nonorthogonal reference manifold, retaining that parent manifold can make
the resulting subspace description more variationally flexible and less
dependent on the quality of the generated states.

The remainder of the paper is organized as follows. Section~\ref{sec:theory}
introduces the dual-space eigenproblem and its principal variational and
conditioning properties. Section~\ref{sec:comp_details} summarizes the state
generators and numerical conventions used in the present tests.
Section~\ref{sec:results} presents results for H$_2$, H$_4$, H$_6$,
N$_2$, and cyclobutadiene including controlled noise studies. Section~\ref{sec:discussion}
discusses the interpretation of the dual-space construction and its relation
to NOCI and quantum subspace methods, and Sec.~\ref{sec:outlook} outlines
future extensions, including applications to more sophisticated
reference-specific correlated-state generators.

\section{Theory}
\label{sec:theory}

\subsection{Parent and generated nonorthogonal spaces}

Let
\begin{equation}
\mathcal P=\operatorname{span}\{\ket{\Phi_I}\}_{I=1}^{N}
\end{equation}
be a linearly independent reference manifold.  The reference-only NOCI
problem is
\begin{equation}
\mathbf H_{\rm rr}\mathbf{\tilde{a}}_k
=E_k^{\rm ref}\mathbf S_{\rm rr}\mathbf{\tilde{a}}_k \;,
%
\label{eq:noci_reference}
\end{equation}
with $(H_{\rm rr})_{IJ}=\langle\Phi_I|\hat H|\Phi_J\rangle$ and $(S_{\rm rr})_{IJ}=\langle\Phi_I|\Phi_J\rangle$.  The references may be
Hartree--Fock solutions, broken-symmetry determinants, spin-recoupled states,
or other chemically motivated configurations; their construction is separate
from the DS-NOCI formalism itself~\cite{thom2009hartreefock,lee2022localized}.
Correlated states $\ket{\Psi_I^{c}}\;(I=1,\ldots,N)$ are associated with
the reference states in $\mathcal P$,
\begin{equation}
\ket{\Psi_I^{c}}
=
\hat U_I(\boldsymbol\theta_I)\ket{\Phi_I},
\qquad I=1,\ldots,N .
\label{eq:generated_state}
\end{equation}
Here, $\hat U_I(\boldsymbol\theta_I)$ denotes a general reference-specific
correlator. The DS-NOCI construction does not depend on its particular
parametrization or on how the parameters $\boldsymbol\theta_I$ are obtained.
In the numerical calculations reported below, $\hat U_I$ is realized using
a unitary coupled-cluster-type parametrization.
The parameters $\boldsymbol\theta_I$, defining unitary operator $\hat U_I(\boldsymbol\theta_I)$, may be supplied perturbatively or
obtained from an iterative optimization.  
A correlated-only calculation diagonalizes in
$\mathcal Q$,
\begin{equation}
\mathbf H_{\rm cc}\mathbf{\tilde{b}}_k
=E_k^{c}\mathbf S_{\rm cc}\mathbf{\tilde{b}}_k ;,
\label{eq:correlated_only}
\end{equation}
where $(H_{\rm cc})_{IJ}=\langle\Psi_I^{c}|\hat H|\Psi_J^{c}\rangle$ and $(S_{\rm cc})_{IJ}=\langle\Psi_I^{c}|\Psi_J^{c}\rangle$.
For comparisons below we refer to this as the correlated-only or NOCI- or NOQE-like
problems; the exact state-generation details are those specified for each
calculation.

\subsection{The dual-space Rayleigh--Ritz problem}

In the DS-NOCI formalism one uses the variational basis space $\mathcal{V}_{DS}$, which is spanned by the states defining $\mathcal{P}$ and $\mathcal{Q}$ spaces, i.e,
\begin{equation}
\mathcal{V}_{DS} = \operatorname{span}\{ \ket{\Phi_1},\ldots,\ket{\Phi_N},
\ket{\Psi_1^{c}},\ldots ,\ket{\Psi_N^{c}}\}
\end{equation}
or $\mathcal V_{\rm DS}=\mathcal P \cup \mathcal Q$ for short.
In this non-orthogonal basis the state corresponding to the  $k$-th root can be expanded as
\begin{equation}
\ket{\Psi_k^{\rm DS}}
=\sum_I a_I^{(k)}\ket{\Phi_I}
+\sum_I b_I^{(k)}\ket{\Psi_I^{c}}.
\label{eq:ds_wavefunction}
\end{equation}
The presence of $\mathcal{P}$ basis, representing for example static correlation effects, allows to decouple static correlation effects from dynamical ones encoded in space $\mathcal{Q}$. This is a critical feature of the DS-NOCI approach resulting in increased accuracies and resiliency of the resulting energies to the possible errors or uncertainties in describing dynamical correlation effects encoded in the $\mathcal{Q}$ basis set. 
The coefficients $\{ a_I^{(k)}\}_{I=1}^N$ and $\{ b_I^{(k)}\}_{I=1}^N$ and corresponding energies $E^{\rm DS}_k$ are obtained from diagonalizing $2N\times 2N$ non-orthogonal problem
\begin{equation}
\begin{pmatrix}
\mathbf H_{\rm rr} & \mathbf H_{\rm rc}\\
\mathbf H_{\rm cr} & \mathbf H_{\rm cc}
\end{pmatrix}
\begin{pmatrix}\mathbf a_k\\\mathbf b_k\end{pmatrix}
=E_k^{\rm DS}
\begin{pmatrix}
\mathbf S_{\rm rr} & \mathbf S_{\rm rc}\\
\mathbf S_{\rm cr} & \mathbf S_{\rm cc}
\end{pmatrix}
\begin{pmatrix}\mathbf a_k\\\mathbf b_k\end{pmatrix},
\label{eq:ds_gevp}
\end{equation}
where, for example,
$(H_{\rm rc})_{IJ}=\langle\Phi_I|\hat H|\Psi_J^{c}\rangle$ and
$(S_{\rm rc})_{IJ}=\langle\Phi_I|\Psi_J^{c}\rangle$.
Because
\begin{equation}
\mathcal P\subseteq\mathcal V_{\rm DS},
\qquad
\mathcal Q\subseteq\mathcal V_{\rm DS},
\end{equation}
the Rayleigh--Ritz principle gives, for exact matrix elements and after a
consistent removal of linearly dependent metric directions,
\begin{equation}
E_0^{\rm DS}\leq E_0^{\rm ref},
\qquad
E_0^{\rm DS}\leq E_0^{c}.
\label{eq:variational_nesting}
\end{equation}
These inequalities express the basic protection supplied by the dual space:
a correlated sector can enlarge the reference ansatz but cannot remove the
reference Ritz solution from the exact variational problem.  They do not imply
that noisy measured matrices remain variational.

\subsection{Reference-space invariance}

The dual-space construction is invariant to components of the generated
states that lie within the reference manifold. Collecting the reference and
generated states into $\boldsymbol\Phi$ and $\boldsymbol\Psi$,
\begin{equation}
\boldsymbol\Psi'=\boldsymbol\Psi+\boldsymbol\Phi\mathbf A
\label{eq:reference_invariance}
\end{equation}
leaves the combined variational space unchanged for any coefficient matrix
$\mathbf A$. Consequently, only the components of the generated states outside
the reference manifold provide independent correlation directions,
\begin{equation}
\boldsymbol Q=(1-\hat P_{\rm ref})\boldsymbol\Psi,
\qquad
\hat P_{\rm ref}
=\boldsymbol\Phi\mathbf S_{\rm rr}^{-1}\boldsymbol\Phi^\dagger.
\label{eq:projected_q}
\end{equation}

Equivalently, defining
\begin{equation}
\mathbf X=\mathbf S_{\rm rr}^{-1}\mathbf S_{\rm rc},
\qquad
\boldsymbol Q=\boldsymbol\Psi-\boldsymbol\Phi\mathbf X,
\label{eq:projection_coeff}
\end{equation}
gives a vanishing reference--$Q$ overlap and
\begin{equation}
\mathbf S_{\rm QQ}
=\mathbf S_{\rm cc}
-\mathbf S_{\rm cr}\mathbf S_{\rm rr}^{-1}\mathbf S_{\rm rc}.
\label{eq:schur_metric}
\end{equation}
Thus, reference-space components of an incomplete correlated state are
variationally redundant, while the independent directions outside
$\mathcal P$ determine the additional correlation recovered by DS-NOCI.
The projection can be applied classically to the Hamiltonian and overlap
matrices. Residual near-linear dependencies are removed by canonical
orthogonalization~\cite{lowdin1950nonorthogonality}, which is particularly
important for noisy quantum subspace diagonalization~\cite{epperly2022qsd}.


\subsection{Size-consistency of DS-NOCI energies}

In general, an arbitrary choice of functions spanning the dual space does not guarantee size consistency of the energy, i.e., the additive separation of the total energy ($E$) into the energies ($E_A$) and ($E_B$) of two noninteracting subsystems ($A$) and ($B$). Size consistency is, however, recovered in the special case in which all functions spanning $\mathcal{V}_{\mathrm{DS}}$ become multiplicatively separable in the noninteracting subsystem limit (NSL). In this limit, the corresponding subsystem basis functions are given by
\begin{eqnarray}
    |\Phi_I\rangle &\xrightarrow{\text{NSL}} & |\Phi_I^{(A)} \Phi_I^{(B)}\rangle \;\; (I=1,\ldots,N)\;, \label{sep1} \\
    |\Psi_I^c\rangle &\xrightarrow{\text{NSL}} & |\Psi_I^{c(A)} \Psi_I^{c(B)}\rangle \;\; (I=1,\ldots,N)\;, \label{sep2}   
\end{eqnarray}

Following Ref.~\cite{lee2022localized}, one can show that the multiplicative separability of the basis functions [Eqs.~(\ref{sep1}) and (\ref{sep2})] implies the corresponding multiplicative separability of the overlap matrix ($\mathbf{S}$). Specifically, in the NSL,
\begin{equation}
\mathbf{S}
\xrightarrow{\mathrm{NSL}}
\mathbf{S}^{(A)}\otimes\mathbf{S}^{(B)},
\end{equation}
where $\mathbf{S}^{(A)}$ and $\mathbf{S}^{(B)}$ are the overlap matrices associated with the functions spanning the subsystem dual spaces $\mathcal{V}_{\mathrm{DS}}^{(A)}$ and $\mathcal{V}_{\mathrm{DS}}^{(B)}$, respectively.

This factorization plays a central role in establishing the size consistency of the resulting energy. Combined with the additive separability of the Hamiltonian matrix $\mathbf{H}$,
\begin{equation}
\mathbf{H}
\xrightarrow{\mathrm{NSL}}
\mathbf{H}^{(A)}\otimes\mathbf{I}^{(B)}
+
\mathbf{I}^{(A)}\otimes\mathbf{H}^{(B)},
\end{equation}
where $\mathbf{I}^{(A)}$ and $\mathbf{I}^{(B)}$ denote the identity matrices in the corresponding subsystem spaces and $\mathbf{H}^{(A)}$ and $\mathbf{H}^{(B)}$ are the Hamiltonian matrices corresponding to subsystems,  it leads directly to the additive separability of the energy,
\begin{equation}
E
\xrightarrow{\mathrm{NSL}}
E_A+E_B.
\label{esep}
\end{equation}
Thus, multiplicative separability of the dual-space basis provides a sufficient condition for recovering size consistency in the NSL. The proof follows closely the arguments used to establish the size consistency of NOCI energies in Ref.~\cite{lee2022localized}.

\subsection{Measurement structure}

The correlated--correlated blocks $\mathbf H_{\rm cc}$ and
$\mathbf S_{\rm cc}$ are already required by the correlated-only
calculation (NOCI, NOQE).  DS-NOCI additionally requires reference--reference and
reference--correlated blocks.  For determinantal references, the
reference--reference quantities can often be evaluated classically using
standard NOCI machinery.  A reference--correlated transition contains one
correlated state rather than two, although its precise circuit and sampling
cost depends on the matrix-element estimator.  The dual-space enlargement
therefore changes the prefactor, but not the quadratic scaling in the number
of retained states, and leaves the original correlated--correlated block
unchanged.

\section{Computational details}
\label{sec:comp_details}

All numerical calculations use a doubles-only unitary coupled-cluster
parametrization for the correlated states,
\begin{equation}
\ket{\Psi_I^c}
=
e^{\hat T_{2,I}-\hat T_{2,I}^\dagger}\ket{\Phi_I},
\end{equation}
with
\begin{equation}
\hat T_{2,I}
=
\frac{1}{4}
\sum_{ijab}
t_{ij}^{ab}(I)
\hat a_a^\dagger
\hat a_b^\dagger
\hat a_j
\hat a_i .
\end{equation}

Unless otherwise specified, the doubles amplitudes are obtained from
reference-specific MP2-like expressions,
\begin{equation}
t_{ij}^{ab}(I)
=
\frac{\langle ab\Vert ij\rangle_I}
{\epsilon_i^{(I)}+\epsilon_j^{(I)}
-\epsilon_a^{(I)}-\epsilon_b^{(I)}} .
\label{eq:mp2_amp}
\end{equation}
The amplitudes are used directly in the unitary parametrization without
subsequent optimization. Regularized or level-shifted variants are specified
for the individual benchmarks where used.

\section{Results}
\label{sec:results}
The numerical results are organized to examine three aspects of the
dual-space construction. We first establish the variational effect of retaining
the reference manifold using complementary reference representations and
increasingly rich multireference spaces. We then examine a chemically relevant
multireference barrier, followed by controlled tests of resilience to imperfect
correlated states and noisy quantum-derived matrix elements.

\subsection{H$_2$: complementary reference representations}

The dissociation of H$_2$ is a useful first test because its apparent
multireference character depends strongly on the orbital representation
\cite{izsak2023electronCorrelation}.  We therefore consider two complementary
reference manifolds and ask the same question in each case: what is gained by
retaining the parent references ($\mathcal{P}$ space) explicitly after correlated companions ($\mathcal{Q}$ space) have
been generated?

\subsubsection{Localized broken-symmetry references}

The first representation follows the construction examined in the original
NOQE work and uses two spin-complementary broken-symmetry UHF references,
\begin{equation}
\ket{\Phi_{A\bar B}}
=
\ket{\phi_A\alpha\,\phi_B\beta},
\qquad
\ket{\Phi_{\bar A B}}
=
\ket{\phi_A\beta\,\phi_B\alpha},
\label{eq:h2_local_refs}
\end{equation}
where $\phi_A$ and $\phi_B$ denote orbitals predominantly localized on the
two hydrogen centres.  The references describe opposite assignments of the
$\alpha$ and $\beta$ electrons.  As the bond is stretched, they approach the
two equivalent atom-localized covalent spin arrangements needed for a
balanced description of the singlet state.
A correlated companion is constructed independently for each UHF reference
using the unregularized opposite-spin UMP2 doubles amplitudes evaluated in
that reference's own orbital basis.  Each resulting unitary state is subsequently
transformed to a common orbital basis before the nonorthogonal matrix elements
are evaluated. 

\begin{figure}
    \centering
    \includegraphics[width=\linewidth]{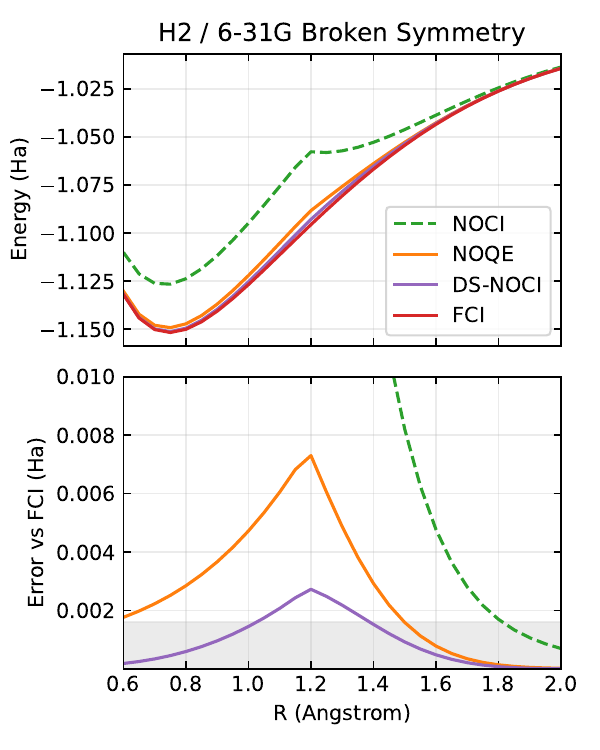}
    \caption{H$_2$/6-31G potential-energy surface obtained from two
    spin-complementary broken-symmetry UHF references.  The upper panel shows
    total energies and the lower panel shows errors relative to FCI.
    Correlated companions are generated using unregularized UMP2
    opposite-spin doubles amplitudes.}
    \label{fig:h2_local_bs}
\end{figure}

The results are shown in Fig.~\ref{fig:h2_local_bs}. At large H--H
separations, the two localized broken-symmetry references approach distinct
UHF dissociation solutions, and the residual dynamic correlation becomes
small. Consequently, reference-only NOCI is most accurate in the dissociation
limit but deteriorates rapidly as the bond is compressed toward the
equilibrium region. NOQE improves substantially upon NOCI at shorter bond
lengths by incorporating dynamic correlation through the correlated sector.
Its largest accuracy is nevertheless obtained in the stretched-bond regime,
and the deviation from FCI remains greater than the chemical-accuracy
threshold over a significant portion of the shorter-bond region. In contrast,
DS-NOCI couples the reference and correlated sectors within a single
variational space and yields energies below those obtained from either NOCI
or NOQE throughout the dissociation coordinate. The resulting error remains
within chemical accuracy over most of the curve, with the principal deviation
occurring in the vicinity of the Coulson--Fischer point, at approximately
$R_{\mathrm{H-H}}\simeq 1.2$~\AA.

\subsubsection{Canonical bonding--antibonding references}

In the canonical molecular-orbital representation, the H$_2$/6-31G
calculation retains all four spatial orbitals,
\begin{equation}
1\sigma_g,\quad 1\sigma_u,\quad 2\sigma_g,\quad 2\sigma_u.
\end{equation}
Their ordering is tracked by symmetry and orbital overlap throughout the
bond-length scan.  We use the two closed-shell references
\begin{equation}
\ket{\Phi_g}=\ket{(1\sigma_g)^2},
\qquad
\ket{\Phi_u}=\ket{(1\sigma_u)^2}.
\label{eq:h2_canonical_refs}
\end{equation}
The bonding configuration dominates near equilibrium, whereas the
antibonding configuration is required to recover the spin-adapted
dissociation limit in a canonical orbital representation.

A correlated companion is generated independently from each reference using
doubles-only $\kappa^2$-regularized MP2-like amplitudes with
$\kappa=0.3\,E_h^{-2}$ \cite{lee2018regularized}.  Here, $\kappa^2$ denotes
the $p=2$ member of the Lee--Head--Gordon $\kappa^p$ regularizer family: the
orbital-energy gap is squared in the exponential damping factor, while the
parameter $\kappa$ itself is not squared.  The amplitudes are used directly,
without singles contributions or subsequent optimization.  For the
non-Aufbau $(1\sigma_u)^2$ determinant, the term MP2-like emphasizes that
this reference is not the stationary RHF solution.

\begin{figure}
    \centering
    \includegraphics[width=\linewidth]{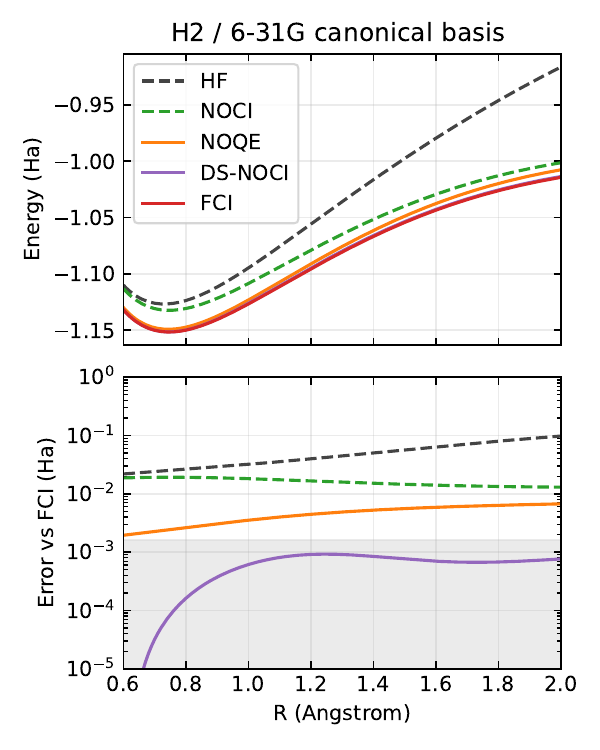}
    \caption{H$_2$/6-31G potential-energy surface in the canonical
    molecular-orbital representation using $(1\sigma_g)^2$ and
    $(1\sigma_u)^2$ as references.  The upper panel shows total energies and
    the lower panel shows errors relative to FCI.}
    \label{fig:h2_631g_pes}
\end{figure}

As shown in Fig.~\ref{fig:h2_631g_pes}, reference-only NOCI captures the
principal bonding--antibonding static correlation but lacks quantitative
dynamic correlation.  Correlated-only NOQE improves the individual parent
states, but the reference and correlation contributions remain combined
within each prepared state.  DS-NOCI instead retains the two bare
configurations and their two correlated companions as independent
variational directions.  The resulting four-state $\mathcal{V}_{\rm DS}$ space improves upon both
component spaces throughout the bond stretch.  Its maximum error relative to
FCI is $0.924$ m$E_h$, so the DS-NOCI curve remains within chemical accuracy
over the complete scan.

The localized broken-symmetry and canonical closed-shell calculations employ
different parent manifolds and different amplitude prescriptions.  Nevertheless,
they exhibit the same variational behavior: the correlated companions are
most effective when they supplement, rather than replace, the parent
reference space.
Hence the dual space setting gives a variational minima and improves upon the parent classical NOCI and the quantum version. 
This becomes more important especially when the amplitudes are not optimized as seen in these cases.

\subsection{N$_2$: enlargement of the seniority-zero reference space}

We next consider N$_2$/STO-6G in a $(6e,6o)$ active space with the active
orbitals ordered as
\begin{equation}
(\pi_y,\pi_x,\sigma_g,\pi_x^\ast,\pi_y^\ast,\sigma_u^\ast).
\end{equation}
For each reference, a correlated companion is generated using doubles-only
level-shifted MP2 amplitudes with a denominator shift of
$\delta=0.2\,E_h$. No subsequent amplitude optimization is performed.

Two closed-shell seniority-zero reference spaces are compared. The
8-reference space contains every configuration obtained by selecting one
doubly occupied orbital from each bonding--antibonding pair,
\begin{equation}
\{\pi_y,\pi_y^\ast\}\times
\{\pi_x,\pi_x^\ast\}\times
\{\sigma_g,\sigma_u^\ast\}.
\end{equation}
To construct the 16-reference space, the remaining 12 seniority-zero
determinants are ranked according to the magnitude of their
Epstein--Nesbet second-order contribution to the 8-reference state, averaged
over the potential-energy scan. Symmetry-related $\pi_x/\pi_y$ partners are
selected together, and the four most important pairs provide the eight
additional references.

For each reference space, we compare reference-only NOCI, correlated-only
NOQE, and DS-NOCI. With eight references, DS-NOCI improves upon both parent
spaces but its error increases during dissociation, reaching approximately
$92\,mE_h$ at $R=2.5$~\AA. Enlarging the parent space to 16 references has
only a modest effect on NOCI or NOQE individually. In particular, at
$R=2.5$~\AA, their errors are approximately $114\,mE_h$ and $233\,mE_h$,
respectively. Their combination in DS-NOCI, however, produces a qualitatively
different result. The maximum DS-NOCI error is $0.313\,mE_h$ at
$R=0.9$~\AA, and from $R=1.0$ to $2.5$~\AA\ the error is below
$10^{-7}\,E_h$.

For the 16-reference calculation, the combined reference and correlated
states span the relevant 32-dimensional symmetry sector of this small active
space. The resulting agreement with FCI is therefore a consequence of the
variational span generated by the dual space, rather than the accuracy of
either parent space individually. This example illustrates the central role
of retaining the bare references: the correlated-only states become
increasingly inaccurate upon stretching, whereas their combination with the
reference manifold recovers the missing variational directions.

\begin{figure*}
    \centering
    \includegraphics[width=\linewidth]{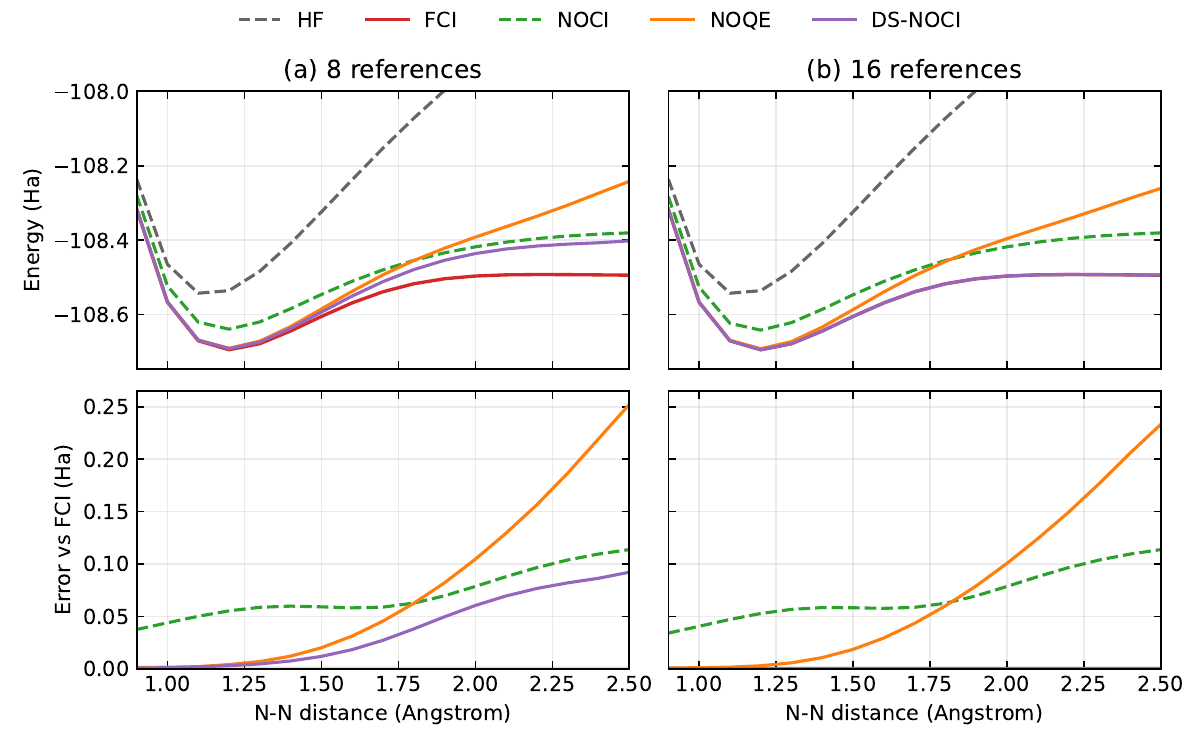}
    \caption{
    N$_2$/STO-6G potential-energy curves in a $(6e,6o)$ active space using
    (a) 8 and (b) 16 closed-shell seniority-zero references. The upper panels
    show the total energies and the lower panels show errors relative to FCI.
    Correlated companions are generated using doubles-only level-shifted MP2
    amplitudes with $\delta=0.2\,E_h$ and no subsequent optimization.
    In panel (b), the DS-NOCI curve is indistinguishable from FCI: its error is
    below $10^{-7}\,E_h$ from $R=1.0$ to $2.5$~\AA, and consequently lies on
    the horizontal axis in the lower panel.
    }
    \label{fig:n2_sto6g_reference_comparison}
\end{figure*}


\subsection{Cyclobutadiene automerization barrier}

Cyclobutadiene automerization is a stringent multireference benchmark because
the electronic structure changes substantially between the rectangular
$D_{2h}$ minimum and the square $D_{4h}$ transition structure. In particular,
the $D_{4h}$ structure exhibits a degenerate frontier-orbital pair and
pronounced multiconfigurational character, requiring a balanced treatment of
static and dynamical correlation
\cite{BalkovaBartlett1994,EckertMaksic2006,Monino2022,ParameswaranTruhlar2026}.

We consider the $D_{2h}\rightarrow D_{4h}$ automerization barrier using the
cc-pVDZ basis and a $(4e,4o)$ active space. The active space contains the
complete carbon $\pi$ manifold, analogous to the six-orbital $\pi$ manifold
of benzene. At the parent $D_{4h}$ geometry, these orbitals transform as
$a_{2u}\oplus e_g\oplus b_{1u}$. The Cartesian geometries are taken from
Ref.~50.

Four closed-shell references are constructed by distributing two electron
pairs among these orbitals:
$\pi(a_{2u})^2\pi(e_g^{(1)})^2$,
$\pi(a_{2u})^2\pi(e_g^{(2)})^2$,
$\pi(e_g^{(1)})^2\pi(b_{1u})^2$, and
$\pi(e_g^{(2)})^2\pi(b_{1u})^2$.
One correlated companion is generated from each reference using
unregularized, reference-specific MP2-based amplitudes, with no level shift
or subsequent amplitude optimization.

\begin{figure}[t]
    \centering
    \includegraphics[width=\linewidth]
    {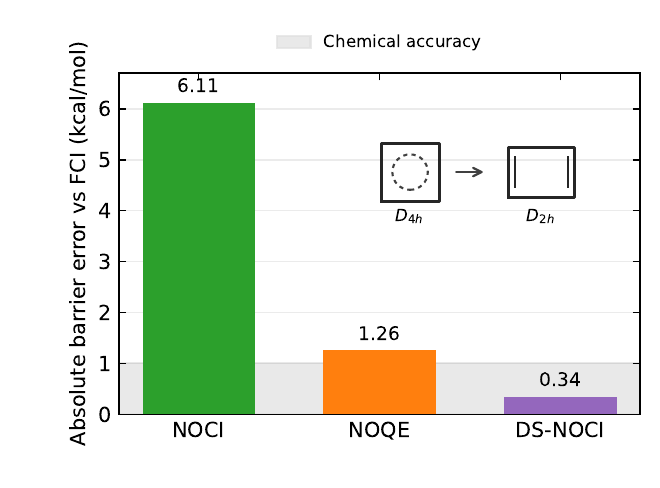}
    \caption{
    Absolute cyclobutadiene automerization-barrier errors relative to
    CASCI/FCI$(4e,4o)$. Correlated companions use unregularized,
    reference-specific MP2-based amplitudes without subsequent optimization.
    The shaded region denotes chemical accuracy.
    }
    \label{fig:cbd_mp2_barrier_error}
\end{figure}

The CASCI/FCI$(4e,4o)$ automerization barrier is $8.96$~kcal/mol.
The corresponding NOCI, NOQE, and DS-NOCI barriers are $15.08$, $7.70$,
and $8.62$~kcal/mol, respectively. Thus, the absolute barrier error is reduced
from $6.11$~kcal/mol for NOCI and $1.26$~kcal/mol for NOQE to
$0.34$~kcal/mol for DS-NOCI (see Fig. \ref{fig:cbd_mp2_barrier_error}). The result demonstrates that the reference and
correlated sectors provide complementary information: although the fixed
MP2-generated states do not reproduce the active-space benchmark on their
own, their coupling to the retained reference manifold yields a substantially
more accurate barrier without any amplitude optimization.

\subsection{Resilience to imperfect correlation models}

The correlated states entering quantum subspace methods may be generated using
fixed perturbative amplitudes or optimized iteratively, for example through
variational quantum eigensolver (VQE) procedures. In either case, the resulting
parameters will generally be approximate because of an incomplete ansatz,
imperfect optimization, or noise in the optimization procedure. DS-NOCI is
agnostic to how these parameters are obtained: the correlated states enter only
as additional variational directions, while the original reference manifold is
retained explicitly. Consequently, inaccuracies in the correlation model do not
force the calculation to rely exclusively on an imperfect correlated space.

\begin{figure}[h]
    \centering
    \includegraphics[width=\linewidth]{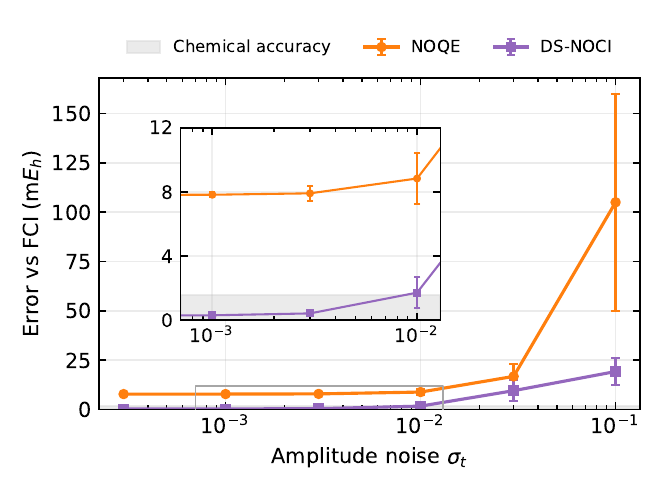}
    \caption{Error relative to FCI for the correlated space only NOQE and DS-NOCI
    descriptions of linear H$_4$/STO-3G at $R=2.0$ bohr as a function of
    Gaussian perturbations to the MP2 doubles amplitudes. Markers denote the
    mean over 200 realizations and error bars indicate the standard deviation.}
    \label{fig:h4_amp_noise}
\end{figure}

We first examine this behavior for linear H$_4$/STO-3G at $R=2.0$ bohr by
perturbing each MP2 doubles amplitude with an independent Gaussian random
variable,
\begin{equation}
t_{ij}^{ab}\rightarrow t_{ij}^{ab}+\eta_{ij}^{ab},
\qquad
\eta_{ij}^{ab}\sim\mathcal N(0,\sigma_t^2),
\end{equation}
where $\mathcal N(0,\sigma_t^2)$ denotes a normal distribution with zero mean
and variance $\sigma_t^2$, and $\sigma_t$ therefore controls the magnitude of
the amplitude perturbation. For each value of $\sigma_t$, 200 independent
realizations of the complete perturbed amplitude set are generated and the
resulting energies are averaged. As shown in Fig.~\ref{fig:h4_amp_noise}, the correlated-only solution becomes
progressively more sensitive to errors in the amplitudes, whereas DS-NOCI
exhibits a substantially weaker dependence over the same range. This behavior
is consistent with the reference-space invariance discussed above: components
of the perturbed correlated states that remain within the retained reference
manifold are variationally redundant, while only changes to the independent
directions outside that manifold modify the dual-space variational space.


A more stringent test is obtained by removing the chemically motivated amplitudes altogether. For H$_6$, we use six low-energy closed-shell references and compare energies obtained using correlated states generated from reference-specific MP2 amplitudes with those obtained using correlated states generated from amplitudes drawn randomly 
from $[-1,1]$. No subsequent amplitude optimization was performed in the latter case.

As shown in Table~\ref{tab:h6_stress}, the correlated-only NOQE description
fails for the random initialization. DS-NOCI remains close to reference-only
NOCI because the bare reference manifold is retained explicitly. With MP2
amplitudes, the correlated states provide physically useful directions, and
DS-NOCI gives the lowest approximate energy.

\begin{table}[t]
\caption{
H$_6$/STO-3G total energies obtained using six low-energy closed-shell
references. The random column corresponds to one reproducible realization
with seed zero. Energies are in hartree.
}
\label{tab:h6_stress}
\begin{ruledtabular}
\begin{tabular}{lcc}
Method & Random amplitudes & Fixed MP2 amplitudes \\
\hline
NOQE    & $-1.28438651$ & $-3.20437173$ \\
DS-NOCI & $-3.13986795$ & $-3.21217519$ \\
\hline
RHF  & \multicolumn{2}{c}{$-3.10585013$} \\
NOCI & \multicolumn{2}{c}{$-3.13844287$} \\
FCI  & \multicolumn{2}{c}{$-3.21769929$} \\
\end{tabular}
\end{ruledtabular}
\end{table}

These tests illustrate that the dual-space construction does not depend on a
particular amplitude model or optimization procedure. When the generated
states are accurate, they provide additional correlation beyond the reference
manifold; when they are poor, the retained reference sector provides a
variational fallback. With exact matrix elements, this inclusion guarantees
that the DS-NOCI ground-state energy is no higher than the corresponding
reference-only NOCI energy.

\subsection{Resilience to matrix-element noise}

In addition to errors in the parameters used to generate the correlated
states, quantum subspace calculations are affected by uncertainty in the
measured Hamiltonian and overlap matrix elements. We examine this source of
error using the two-reference H$_2$/6-31G calculation by perturbing the exact
subspace matrices according to
\begin{equation}
\mathbf H'=\mathbf H+\Delta\mathbf H,
\qquad
\mathbf S'=\mathbf S+\Delta\mathbf S,
\end{equation}
where $\Delta\mathbf H$ and $\Delta\mathbf S$ represent statistical
perturbations to the measured matrix elements.

The reference--reference blocks are assumed to be evaluated classically
(or the errors in evaluating the corresponding matrix elements are assumed to be negligible) 
and are therefore kept exact,
\begin{equation}
\Delta\mathbf H_{\rm rr}=0,
\qquad
\Delta\mathbf S_{\rm rr}=0.
\end{equation}
Matrix elements involving the correlated sector are perturbed independently
with zero-mean Gaussian noise. For the correlated--correlated block,
\begin{equation}
(\Delta H_{\rm cc})_{IJ}
\sim\mathcal N(0,\sigma_H^2),
\qquad
(\Delta S_{\rm cc})_{IJ}
\sim\mathcal N(0,\sigma_S^2),
\end{equation}
where $\sigma_H$ and $\sigma_S$ set the standard deviations of the
Hamiltonian and overlap perturbations, respectively. In the present test,
\begin{equation}
\sigma_H=0.012\,E_h,
\qquad
\sigma_S=0.004.
\end{equation}
The perturbed matrices are constructed to preserve Hermiticity.

The reference--correlated block is treated separately. These matrix elements
involve only one correlated state preparation, in contrast to the
correlated--correlated block, for which both states are generated by
correlated circuits. To represent a lower-noise cross block in the present
numerical model, we therefore use
\begin{equation}
\sigma_H^{\rm rc}=\frac{\sigma_H}{\sqrt{2}},
\qquad
\sigma_S^{\rm rc}=\frac{\sigma_S}{\sqrt{2}},
\end{equation}
such that
\begin{equation}
(\Delta H_{\rm rc})_{IJ}
\sim\mathcal N\!\left(0,\frac{\sigma_H^2}{2}\right),
\qquad
(\Delta S_{\rm rc})_{IJ}
\sim\mathcal N\!\left(0,\frac{\sigma_S^2}{2}\right).
\end{equation}
The corresponding correlated-reference blocks are fixed by Hermiticity. The factor of
$1/\sqrt{2}$ is a modeling choice used to distinguish the two classes of
matrix elements and should not be interpreted as a universal relation between
their measurement uncertainties.

Because noise in the overlap matrix can introduce near-linear dependencies
or amplify pre-existing ones, each noisy generalized eigenvalue problem is
solved following canonical orthogonalization of the metric. We also consider
the projected representation introduced in
Eq.~\eqref{eq:projection_coeff}, in which components of the correlated states
within the reference manifold are removed by a classical basis
transformation before diagonalization. No additional quantum-state
preparation is required for this projection.
Hence the dual space also helps in a way to mitigate the errors at the classical matrix element level and does not introduce any expensive projection in the quantum hardware. 

\begin{figure}
\centering
\includegraphics[width=0.95\linewidth]{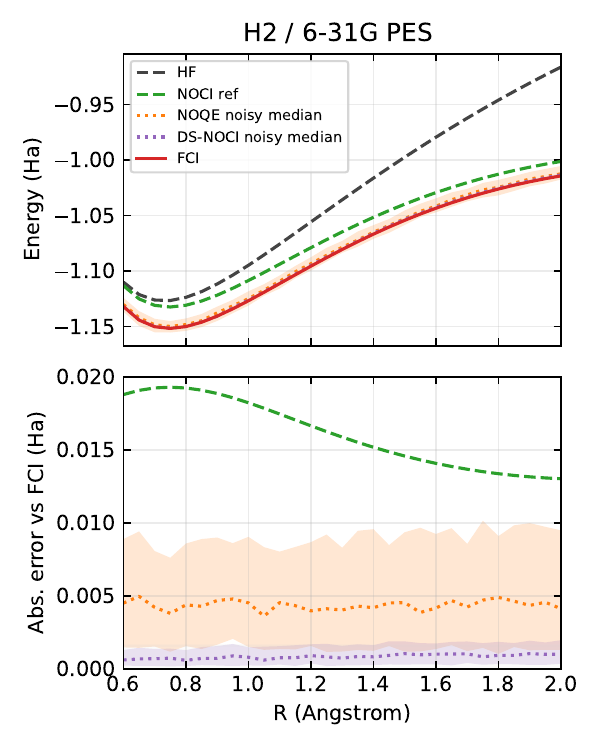}
\caption{Sensitivity of the H$_2$/6-31G two-reference calculation to
statistical perturbations of the subspace Hamiltonian and overlap matrices.
Curves show the median error relative to FCI over
100 independent noisy runs, and shaded regions denote the
16th--84th percentile interval.}
\label{fig:h2_noise}
\end{figure}

For each bond length, 100 independent realizations of the noisy Hamiltonian
and overlap matrices are generated. Figure~\ref{fig:h2_noise} reports the
median error relative to FCI, with the 16th--84th percentile interval used to
characterize the spread of the resulting energy distribution.

As shown in Fig.~\ref{fig:h2_noise}, DS-NOCI remains closer to FCI than the
correlated-only calculation over the reported dissociation coordinate and
exhibits a narrower central error distribution. The improvement originates
from the retained reference sector, whose matrix elements are unaffected by
the simulated quantum measurement noise, together with the variational
coupling to the noisy correlated sector. The dual-space construction therefore reduces the sensitivity of the final
energy to errors in the quantum-derived matrix elements, although it does not
eliminate their effect. As the uncertainty in the correlated and cross blocks
increases, the information carried by the generated states becomes
progressively less reliable, and the advantage of enlarging the variational
space correspondingly decreases. This fundamental limitation is not removed
at the DS-NOCI level. However, established hardware-level error-mitigation strategies, including
measurement-error mitigation~\cite{bravyi2021measurement},
symmetry verification or postselection~\cite{sagastizabal2019symmetry},
and zero-noise extrapolation~\cite{temme2017error}, can be applied to the
expectation values entering the DS-NOCI Hamiltonian and overlap matrices and
may therefore provide further improvements in the resulting energies.

\section{Discussion}
\label{sec:discussion}

The defining feature of DS-NOCI is the explicit coexistence of two spaces:
a parent nonorthogonal reference manifold and a manifold of correlated states
generated from those references.  This should be distinguished from the
particular choice of orbitals, amplitudes, or quantum circuit used to produce
the latter.  The numerical examples use inexpensive state generators because
they expose the effect of the dual-space construction cleanly; the theory does
not require them to remain fixed or perturbative.

This distinction also clarifies the relation to existing quantum subspace
methods.  Krylov, filter-diagonalization, QSE, NOVQE, NOQE, and GCIM already
show that projected diagonalization in a compact nonorthogonal span can be a
powerful alternative to solving the full Hilbert-space problem
\cite{mcclean2017hybrid,motta2020qite,stair2020mrqk,cortes2022quantum,parrish2019qfd,huggins2020nonorthogonal,baek2023noqe,zheng2024gcim}.
DS-NOCI's specific construction preserves the \emph{parent (reference)  NOCI manifold} as an
independent variational sector after reference-specific correlated quantum
states have been generated, and solves the union of the two spaces.

The reference-space invariance in Eq.~\eqref{eq:reference_invariance} gives a
simple way to understand why this can help.  A generated state may contain a
large reference-like component.  In a correlated-only basis that component is
locked to the external correlation direction by the state-preparation
parameters.  In the dual space the reference component is already available
independently, so only the direction outside the parent manifold carries new
variational content.  This is why amplitude errors need not translate one for
one into errors in the DS variational span.  It is also why the projected
Schur metric in Eq.~\eqref{eq:schur_metric} is a useful diagnostic: its small
eigenvalues identify generated combinations that add little independent
information and are especially vulnerable to noise.

The same reasoning explains what DS-NOCI does \emph{not} guarantee.  If the
parent reference manifold is qualitatively inadequate, retaining it cannot
supply missing correlation channels by itself.  If the generated states add
no independent directions, the DS energy simply reduces to a parent-space
result.  If measured matrix elements are noisy, the generalized eigenproblem
is no longer strictly variational and metric regularization becomes essential.
The advantage is therefore structural rather than absolute: the generated
states are allowed to improve a reliable parent description without being
forced to replace it.

There is also no conceptual conflict between DS-NOCI and optimization.  One
may use fixed perturbative amplitudes when low cost is the priority, as in many
of the tests here, or optimize the generated states when additional quantum or
classical resources are justified. Likewise, the formalism can be used with fixed orbitals or with
reference states produced by an orbital-optimization procedure.  Neither
choice changes the definition of the dual-space eigenproblem.

\section{Outlook}
\label{sec:outlook}

The favorable accuracy of DS-NOCI, which outperforms the other NOCI-type approaches considered in this work, particularly in the presence of strong correlation effects, together with the guaranteed size consistency of the resulting energies (assuming separability of all basis functions), makes it a promising framework for simulating strongly correlated systems on both classical and quantum computing platforms.
An additional attractive feature of DS-NOCI is its resilience to errors and uncertainties in the description of correlation effects incorporated into the $|\Psi_I^c\rangle$ basis. This property opens the possibility of developing a two-tier strategy for error suppression in quantum simulations, in which errors are controlled both at the level of the many-body formulation and through error-mitigation or error-correction techniques applied to the quantum circuits used to evaluate the Hamiltonian  and overlap matrices.

DS-NOCI also provides a natural connection to the recently developed parallel quantum flow (QFlow) framework
\cite{kowalski2023qflow,kowalski2025resource}, which is designed to capture correlation effects in large target spaces using reduced quantum resources and low-depth quantum circuits. Recent QFlow simulations have demonstrated ground-state calculations for target spaces spanning 80--114 orbitals while optimizing more than $10^6$ wave-function parameters \cite{bauman2026parallelqflow}. The quantum-resource requirements can be further reduced by exploiting the observation that only a limited subset of reduced-dimensional active spaces is needed within the self-consistent QFlow procedure to recover the dominant correlation effects.

These features suggest a natural integration of DS-NOCI with a multi-reference-type  QFlow strategy. In particular, high-quality $|\Psi_I^c\rangle$ basis functions could be generated through QFlow calculations employing the corresponding $|\Phi_I\rangle$ states as reference functions. Such a construction would provide an alternative to perturbative estimates of $|\Psi_I^c\rangle$ or approximations based on the unitary cluster Jastrow (UCJ)-type expansion \cite{matsuzawa2020jastrow} or its local variant \cite{motta2023bridging}. More broadly, this strategy would enable the parallel utilization of a large number of distributed, small-scale quantum resources for multi-reference QFlow calculations, while larger quantum processors could be employed for the evaluation of $(H_{\rm cc})_{IJ}$, $(H_{\rm rc})_{IJ}$, $(S_{\rm cc})_{IJ}$, $(S_{\rm rc})_{IJ}$ through Hadamard-test-type circuits. The resulting hierarchical framework would therefore combine quantum resources operating at different scales with error control at both the many-body-theory and quantum-circuit levels. We anticipate that such a two-tier approach to error suppression, coupled with heterogeneous and distributed utilization of quantum resources, could provide a promising pathway toward chemically relevant simulations on early fault-tolerant quantum computing platforms.

\section{Conclusions}

We have introduced Dual-Space Nonorthogonal Configuration Interaction, a quantum-subspace construction that explicitly retains a parent
nonorthogonal reference manifold alongside correlated states generated from
those references.  The novelty is the dual-space variational assembly, not a
requirement that the constituent states be produced without optimization.

With exact matrix elements, the combined space contains both the
reference-only and correlated-only parent spaces, giving a direct
Rayleigh--Ritz bound relative to either.  Because additions to the generated
states from within the retained reference manifold leave the combined span
unchanged, only their independent projected directions contribute new
variational information.  This provides a compact explanation for the
stabilizing behavior observed when the generated states are imperfect.

Proof-of-principle H$_2$, H$_4$, H$_6$, N$_2$, and cyclobutadiene calculations show that the
explicit parent sector can improve energies over correlated-only
diagonalization, remain useful as the correlated states are progressively
optimized, and reduce sensitivity to controlled state-parameter and
matrix-element perturbations.  These results support a modular view of quantum
subspace design: preserve a physically meaningful multireference manifold,
generate additional correlated directions with whatever state-preparation
method is appropriate, and determine their final balance by a single coupled
nonorthogonal diagonalization.

\begin{acknowledgments}
This material is based upon work supported by the ``Embedding QC into Many-body Frameworks for Strongly Correlated Molecular and Materials Systems''  project, which is funded by the U.S. Department of Energy, Office of Science, Office of Basic Energy Sciences, the Division of Chemical Sciences, Geosciences, and Biosciences (under FWP 72689) and by the Quantum Algorithms and Architecture for Domain Science Initiative (QuAADS), under the Laboratory Directed Research and Development (LDRD) Program at Pacific Northwest National Laboratory (PNNL). 
V.A. and B.P. acknowledge the support from the Early Career Research Program by the U.S. Department of Energy, Office of Science, under Grant No. FWP 83466.
This work used resources from the PNNL.
NPB and BJ ackowledge support by the U.S. Department of Energy, Office of Science,
Basic Energy Sciences, Division of Materials Sciences and
Engineering, Theoretical Condensed Matter Physics Program, FWP 83557.
PNNL is operated by Battelle for the U.S. Department of Energy under Contract DE-AC05-76RL01830.
\end{acknowledgments}

\appendix
\section{Nonorthogonal L\"owdin--Feshbach elimination}
\label{app:downfold}

For interpretation, the correlated sector of Eq.~\eqref{eq:ds_gevp} can be
eliminated exactly at a trial energy $E$. The block equations are
\begin{align}
(\mathbf H_{\rm rr}-E\mathbf S_{\rm rr})\mathbf a
+(\mathbf H_{\rm rc}-E\mathbf S_{\rm rc})\mathbf b&=0,\\
(\mathbf H_{\rm cr}-E\mathbf S_{\rm cr})\mathbf a
+(\mathbf H_{\rm cc}-E\mathbf S_{\rm cc})\mathbf b&=0.
\end{align}
When $E\mathbf S_{\rm cc}-\mathbf H_{\rm cc}$ is invertible,
\begin{equation}
\mathbf b
=(E\mathbf S_{\rm cc}-\mathbf H_{\rm cc})^{-1}
(\mathbf H_{\rm cr}-E\mathbf S_{\rm cr})\mathbf a,
\end{equation}
which gives
\begin{equation}
[\mathbf H_{\rm rr}+\boldsymbol\Sigma_{\rm rr}(E)]\mathbf a
=E\mathbf S_{\rm rr}\mathbf a,
\end{equation}
with
\begin{equation}
\boldsymbol\Sigma_{\rm rr}(E)=
(\mathbf H_{\rm rc}-E\mathbf S_{\rm rc})
(E\mathbf S_{\rm cc}-\mathbf H_{\rm cc})^{-1}
(\mathbf H_{\rm cr}-E\mathbf S_{\rm cr}).
\label{eq:self_energy_appendix}
\end{equation}
Thus the generated sector can be viewed as an energy-dependent self-energy
that dresses both diagonal reference energies and off-diagonal reference
mixing. The retained metric remains $\mathbf S_{\rm rr}$; nonorthogonality
between the sectors is carried explicitly by Eq.~\eqref{eq:self_energy_appendix}.
DS-NOCI itself is solved using the full generalized eigenproblem rather than
iterating this reduced equation. An analogous procedure can be used to define the self-energy $\boldsymbol\Sigma_{\rm cc}(E)$ in the correlated–correlated sector of the dual space,
\begin{equation}
[\mathbf H_{\rm cc}+\boldsymbol\Sigma_{\rm cc}(E)]\mathbf b =
E\mathbf S_{\rm cc}\mathbf b, \label{se1} 
\end{equation}
and 
\begin{equation}
\boldsymbol\Sigma_{\rm cc}(E)=
(\mathbf H_{\rm cr}-E\mathbf S_{\rm cr})
(E\mathbf S_{\rm rr}-\mathbf H_{\rm rr})^{-1}
(\mathbf H_{\rm rc}-E\mathbf S_{\rm rc}).
\label{se2}
\end{equation}
The $\boldsymbol\Sigma_{\rm cc}(E)$ describes coupling mechanisms between $\mathcal{Q}$ and $\mathcal{P}$ spaces.

\bibliographystyle{apsrev4-2}
\bibliography{library_revised}

\end{document}